# Developments in Connected Vehicles and the Requirement for Increased Cybersecurity

Author 1
Mr. Phillip Garrad*
Institute of Technology Sligo
phillgarrad@gmail.com
*Corresponding author

Author 2
Mr. Shane Gilroy
Institute of Technology Sligo
gilroy.shane@itsligo.ie

**Abstract**

The increase in popularity of connected features in intelligent transportation systems, has led to a greater risk of cyber-attacks and subsequently, requires a more robust validation of cybersecurity in vehicle design. This article explores three such cyber-attacks and the weaknesses in the connected networks. A review is carried out on current vulnerabilities and key considerations for future vehicle design and validation are highlighted. This article addresses the vehicle manufacture's desire to add unnecessary remote connections without appropriate security analysis and assessment of the risks involved. The modern vehicle is "All Connected" and only as strong as its weakest link.

## Introduction

The growth of connected features in intelligent transport systems has been proportional to the increase of automotive cyber-attacks in recent years [1]. The first consumer connected vehicle was manufactured by General Motors Cadillac in 1996 [2]. The connectivity feature, known as the OnStar system, consisted of an emergency messaging service which contacted authorities with the GPS location when the airbags were deployed, such as at the time of a collision. However, in the event of the system being exploited, a hacker could leverage available data to track the car and trigger fake emergency response calls from the vehicle. In the following years OnStar was enhanced further and by 2003 it was able to perform vehicle health reports and turn-by-turn directions through a connection to the vehicle framework and an external secure network, respectively. When this data is exposed to hackers, user location and other personal data is accessible along with the route plan of the vehicle. Fortunately, these systems are still functionally secure with regards to road safety of the vehicle as the connected system only comprised of advisory messages and navigation instructions. More involved control-by-wire systems such as Tesla's Autopilot [3] in 2014 were released to the public, presenting a greater risk of cyber-attacks affecting road safety.

Vehicle designers are integrating more and more connected features into their products with a view to providing enhanced safety and user experience for their customers [4] leading to an increase in the number of access points which require robust cybersecurity features. This paper describes three case studies of cyber-attacks in recent years and how they were handled. It will then explore the evolution of cybersecurity technology and how these attacks may have been prevented.

## Case Studies

The need for enhanced cybersecurity increases as more connected features are integrated into vehicles. In 2016, the U.S. Department of Transportation and the National Highway Traffic Safety Administration (NHTSA) first published a public warning on the dangers of connected vehicles and the risk of security systems being penetrated by an outsider [5]. The following sections will discuss security system failures of connected vehicles from a variety of manufacturers.

### Jeep, Cherokee – 2015

A 2015 Jeep Cherokee was hacked by Miller and Valasek [6], two cybersecurity experts. Previously, the duo had exploited open cyber-attack vectors by hacking a Ford Escape through the OBD-II (On Board Diagnostics) connection. However, this time the hacking was performed remotely over 15 kilometres away through the internet connection on the vehicle. The Uconnect feature was exploited, which allowed access to the vehicle's infotainment system remotely. The feature contained a vulnerability that allowed the IP address to be accessible to the hacker. The hackers then gained access to drive by wire systems and were able to toggle the ignition, lower the engine speed, abruptly engage brakes or disable the brakes. They were also able to access the Jeep's GPS and IMUs (Inertial Movement Units) to measure its speed, distance, direction. Additionally, the Uconnect feature allowed them to control the heating and radio systems. Once a vehicle's IP address is exposed, the vehicle can be hacked anywhere, anytime.

### Mitsubishi Outlander Plug-in Hybrid Electric Vehicle (PHEV) – 2016

PenTestPartners exploited a 2016 Mitsubishi Outlander PHEV [7]. Many new cars use a Global System for Mobile Communications (GSM) module so the owner may communicate with the vehicle when away from it to track its location, set the heating, open windows etc. The 2016 Mitsubishi Outlander PHEV model uses a Wi-Fi access point instead of a GSM module for connected services. This greatly limits the range the user can remotely access their vehicle and leaves it more susceptible to attacks by using a less common method of remote connection. PenTestPartners took advantage of this vulnerability [7], by cracking the Wi-Fi Pre-Shared Key (PSK), and then only required access the vehicle while it was connected to an authorised phone. This provides access to disable the alarm and even unlock the car. Once they knew a vehicle's SSID (Service Set Identifier) any vehicle of the same make and model could be found and connected to a mobile device in order to determine its geolocation.

### Nissan Leaf – 2016

A similar attack occurred with a 2016 Nissan Leaf in the same year [8]. In this case, Nissan's "NissanConnect" software required the user to enter the vehicle identification number (VIN) into an app to connect to the vehicle. In many cases the VIN was stamped on the windscreen for easy access. As the VIN is composed of the manufacturer code, model code and country code, cybercriminals only needed approximately 5 digits to vary between vehicles. This connection can be established remotely and once connected a hacker can determine the vehicle's geolocation and control all heating and cooling units (air, seat, steering wheel).

### Other Recent Cyber-Attacks

Further research of case studies in recent years has made note of hacks primarily involving eavesdropping of sensitive data and GPS coordinates [9]. In 2018 three of BMW's models were proven to have vulnerabilities through OBD connection allowing the hacker to shut off the vehicle by overloading the system with erroneous messages [10]. In 2017 a Tesla fleet was hacked by exploiting a bug in the fleet's central sever. The hacker gained access to control any car in the fleet, being able to pass Tesla commands, such as drive home remotely [11]. Other cyber-attacks on Tesla in 2018 involved communication jamming by sending noisy messages over the local network [9].

Context Information Security exposed vulnerabilities in the 2020 Ford Focus and Volkswagen Polo [12]. Weaknesses were discovered in the infotainment system of the 2020 Volkswagen Polo which when exploited granted a hacker access to toggle the traction control and review personal data of the driver. Context Information Security was also used to perform a "man in the middle" attack on the Ford Focus, intercepting messages from the TPMS (Tyre Pressure Monitoring System). When presented with the technical report Ford choose to ignore it whereas Volkswagen choose to implement countermeasures.

**Challenges and Countermeasures**

This section outlines the main challenges faced by the vehicle manufacturers and how they can be addressed:

- The Automotive Industry faces similar challenges to the Software Industry but with higher stakes: Due to the complexity and redundancy required to implement a connected vehicle, the average modern high-end car has 100 million lines of code [13]. Facebook, which consists of 51 million lines of code and has 108 listed vulnerabilities in the National Vulnerability Database in the past 3 years [14]. Facebook has also been publicly hacked once so far in 2021 and on three occasions in 2019, exposing confidential information of close to 1.5 billion users [15], [16], [17]. This demonstrates that modern software is also challenging to design with sufficient attention to security.
- Security is an afterthought, or a patch: Bug fixing is a common approach to handling software flaws found in software. When it comes to cyber vulnerabilities, considering the cyber security risks at the design phase is a prerequisite to developing a connected vehicle [18].
- Supply chains which requires thorough validation: In vehicle manufacturing and assembly, there is a long line of suppliers for each component. Cybersecurity must be considered during component integration involving a knowledge share between suppliers [18]. A study showed less than 40% of original equipment manufacturers and suppliers are aware of knowledge sharing for component integration taking place [19].
- Lack of sufficient stress testing: There are insufficient professional white coat hackers. More than half of the cyber-attacks are performed by cyber criminals rather than researchers [20]. To reduce the risk of vehicles being hacked in the field more extensive real world ethical hacking should be done during design and test phases.

**Types of Attacks**

As more connected features are integrated to vehicles, there is a higher risk of vulnerabilities [21]. Connected vehicles are vulnerable to two modes of cyber-attacks, active and passive. An active attack is when communications are disrupted or replaced by falsified messages, a passive attack is where data is collected over a longer time [22]. These attack modes consist of multiple attack types. The attack-type is determined by the method used to carry out the hack. Active attacks include spoofing, dysfunctional sensor processing, side channel attacks. Passive attacks include man-in-the-middle attacks. Active attacks are more dangerous but fortunately easier to detect as the functionality of the vehicle becomes compromised [22]. Passive attacks do not change or interact with the functions of a vehicle, merely listen to the communications.

**Communication layer**

A hacker can gain access to a vehicle through several ways. External communications such as V2X (Vehicle-to-Everything) and infotainment systems are entry points for hackers. Internal access points are OBD-II and USB ports. However, once the hacker has gained access to any of these access points, they may be able to connect to the various internal communications [23]. Figure 1 below shows the common attack types, attack vectors and attack surfaces that are exploited by hackers [21].

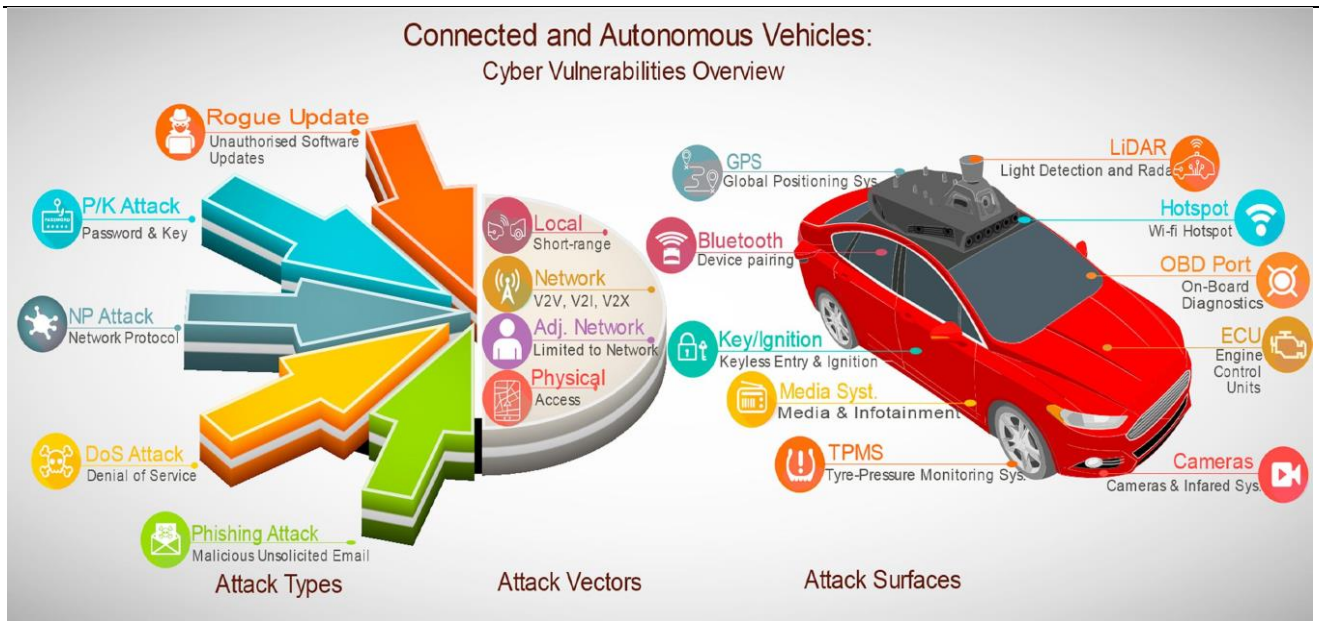

**Figure 1: Overview of cyber-attack types, attack vectors and attack surfaces [21]**

## Attack Surfaces

Each function has its own network protocol based on its requirement [24]. For example, the Media Oriented Systems Transport (MOST) bus systems are used for media and infotainment around the vehicle and it is typically not connect to any of critical systems. Controller Area Network (CAN) connects the powertrain sensors, hybrid drive, transmission and engine controller. It checks the data at each node to verify it is secure and related to that node, if not it skips it. If exposed the CAN bus is far more valuable to a hacker. The attack surface varies as detailed below:

- GPS: GPS can be spoofed by a hacker sending misleading messages to the connected vehicle. Alternatively, the GPS can be hacked using a signal jammer. A signal jammer sends a strong signal repeatedly to prevent the valid signal to be received [25].
- Key/Ignition: The key has always been the prime means of accessing a vehicle. Key fob hacks as mentioned in the case studies in section 2, and with electronic key fob cloning [26] this remains a concern.
- Media System: The media system commonly refers to infotainment and telematics. Infotainment is an umbrella term for in-car information and entertainment. Some of this information is text messages, notifications etc. received from connecting to a mobile phone. It also includes streaming music and videos from a mobile device. Telematics refers to internal vehicular data such as fuel efficiency, vehicle speed, climate control, direction, odometer, among others. Infotainment and Telematics systems are vulnerable to attacks by malware which can be installed through the owner's phone. Countermeasures to prevent these attacks include using source code obfuscation to make it harder for the hacker to create malware [27].
- TPMS: Tyre Pressure Monitoring Systems have a built-in sensor on the wheel which connects to the vehicle using a short-range signal such as Bluetooth [28].
- LiDAR: LiDAR (Light Detection and Ranging) is a key sensor used for perception and localisation of connected vehicles with a high level of automation. As it uses laser light for detection, it can be hacked externally by using strong lights to imitate or hide objects [29].
- Bluetooth and Wi-Fi Hotspot: These are exploited in similar ways as they require the hacker to be in close range with the vehicle to connect to the system. With successful connection to these attack surfaces a hacker can typically access the Media System.
- OBD-Port: OBD-II monitors control and engine systems and can send an error to the vehicle user if a

sensor fails. It is used for vehicle diagnostics and is in all vehicles since 1996 in the US and Ireland. This port can be hacked by installing malware through direct connection. Alternatively, a dongle can be used and the OBD-II may be hacked remotely. Monitoring the frame injection to the OBD-II is a way to prevent this. Also, signing and encrypting of firmware updates should be carried out by the manufacturer [24].

- ECU: ECU's (Electronic Control Units) are connected to the vehicle's CAN network, passing data from the drivetrain and a range of other sensors. The ECUs are typically not connected to a remote network however by passing an erroneous signal through peripheral sensors these too can be exploited [30].
- Cameras: On connected vehicles, cameras are used to monitor the vehicle environment. With automated functions such as ADAS (Advanced Driving Assistance Systems) camera data is processed to control the vehicle and guide it. Cameras can be hacked by passing fake data [29].

**Case Study Countermeasures**

Once vulnerabilities are identified by vehicle manufacturers, security measures can be taken. In the case of the Jeep, Cherokee – 2015, Fiat Chrysler recalled 1.4 million cars and supplied a secure update to prevent this attack from re-occurring in the future [6]. For the Mitsubishi Outlander Plug-in Hybrid Electric Vehicle (PHEV) – 2016 the vehicle manufacturer updated the firmware for the Wi-Fi module to implement a stronger PSK, blocking future occurrences of this cyber-attack, allowing users to reenable the app [7]. For the remainder of that vehicle model Mitsubishi continued to use the Wi-Fi Module over the more secure GSM module for cost purposes but swapped to the GSM Module in 2019 with the newer model. An alternative solution for this issue is to disconnect and unpair all mobile devices from the vehicle the Wi-Fi modules essentially goes to sleep. Lastly, the Nissan Leaf – 2016 hack resulted in Nissan disabling "ConnectEV", while an updated version was developed [8].

**Conclusions**

The rise in uptake in connected functions in modern vehicles has led to increase in the number of cyber-attacks committed in recent years. Connected vehicle technology will continue to progress with the rise of automated and autonomous vehicles. In order to maintain safety and security while increasing the number of connected vehicle functions the automotive industry, academia and government bodies need to be aligned, especially in terms of safety standards. As technology advances, safety standards and regulations should be reviewed and updated as appropriate.

From the case studies outlined, it is evident that further validation and oversight is required in the development cycle of connected vehicle functions. The cases studies and corresponding countermeasures discussed provide examples of vulnerabilities that were discovered only after public release of the vehicles. Each of the weaknesses could have been discovered and corrected in the development cycle of the vehicle with thorough penetration testing. It is important that vehicle manufacturers continue to learn from these novel cybersecurity trends. There are many connected systems such as "Uconnect" and "ConnectEV" that do not directly benefit the vehicle's functionality and could be removed to balance resources and focus cybersecurity on key areas, such as drive-by-wire systems more efficiently. It is important for automotive manufacturers to remember that the "All Connected" vehicle is only as strong as its weakest link.